\documentclass[11pt]{article}
\usepackage{moriond,epsfig}
\usepackage {color}

\def\Journal#1#2#3#4{{#1} {\bf #2}, #3 (#4)}

\def\NIMA{{\em Nucl. Instrum. Methods} A}
\def\NPB{{\em Nucl. Phys.} B}
\def\PLB{{\em Phys. Lett.}  B}
\def\PRL{\em Phys. Rev. Lett.}
\def\PRD{{\em Phys. Rev.} D}

\def\EPJ{{\em Eur. Phys. J.} C}

\def\be{\begin{equation}}
\def\ee{\end{equation}}
\def\bea{\begin{eqnarray}}
\def\eea{\end{eqnarray}}

\def\ff{$\phi$ factory}

\newcommand{\R}{\ensuremath{R_{S}^{\pi}}}
\newcommand{\Dafne}       {DA\char8NE}
\newcommand{\ks}          {\ensuremath{K_{S}}}
\newcommand{\kl}          {\ensuremath{K_{L}}}
\newcommand{\kp}          {\ensuremath{K^{+}}}
\newcommand{\km}          {\ensuremath{K^{-}}}
\newcommand{\kpm}          {\ensuremath{K^{\pm}}}

\newcommand{\Pem}{\ensuremath{e^-}}
\newcommand{\Pep}{\ensuremath{e^+}}

\newcommand{\Pnu}{\ensuremath{\nu}}
\newcommand{\Pnubar}{\ensuremath{\bar{\nu}}}
\newcommand{\Pphi}{\ensuremath{\phi}}

\newcommand{\Ppio}{\ensuremath{\pi^0}}
\newcommand{\Ppim}{\ensuremath{\pi^-}}
\newcommand{\Ppip}{\ensuremath{\pi^+}}

\newcommand{\toP}{\ensuremath{\rightarrow}}
\newcommand{\DKSeIII}{\ensuremath{K_S\rightarrow\pi e \nu}}

\newcommand{\DKSeIIIeppm}{\ensuremath{K_S\rightarrow\Ppim\Pep\Pnu}}
\newcommand{\DKSeIIIempp}{\ensuremath{K_S\rightarrow\Ppip\Pem\Pnubar}}

\newcommand{\KSpippim}{\ensuremath{K_S\rightarrow\pi^+\pi^-}}

\newcommand{\DKLeIII}{\ensuremath{K_L\rightarrow\pi e \nu}}
\newcommand{\DKLmuIII}{\ensuremath{K_L\rightarrow\pi \mu \nu}}
\newcommand{\DKLpiopiopio}{\ensuremath{K_L\rightarrow\pi^0\pi^0\pi^0}}
\newcommand{\DKLpippimpio}{\ensuremath{K_L\rightarrow\pi^+\pi^-\pi^0}}

\newcommand{\Dkmudue}[1]{\ensuremath{K^{{#1}} \to \mu^{{#1}}\nu}}
\newcommand{\Dpimudue}[1]{\ensuremath{\pi^{{#1}} \to \mu^{{#1}}\nu}}
\newcommand{\Dkpidue}[1]{\ensuremath{K^{{#1}} \to \pi^{{#1}}\pi^{0}}}%

\newcommand{\kmudue}[1]{\ensuremath{K^{{#1}}_{\mu2}}}
\newcommand{\kpidue}[1]{\ensuremath{K^{{#1}}_{\pi2}}}
\newcommand{\ketre}[1] {\ensuremath{K^{{#1}}_{e3}}}
\newcommand{\kmutre}[1]{\ensuremath{K^{{#1}}_{\mu3}}}
\newcommand{\kltre}[1] {\ensuremath{K^{{#1}}_{{\rm l}3}}}
\newcommand{\eV}{{e\kern-.07em V}}
\newcommand{\MeV}{{\rm \,M\eV}}

\newcommand{\GeV}{{\rm \,G\eV}}

\newcommand{\ps}{{\rm \,ps}}
\newcommand{\ns}{{\rm \,ns}}
\newcommand{\mm}{{\rm \,mm}}
\newcommand{\m}{{\rm \,m}}
\newcommand{\cm}{{\rm \,cm}}

\newcommand{\Lpb}{\ensuremath{\rm \, pb^{-1}}}
\newcommand{\BR}[1]{\ensuremath{\mathrm{BR}(#1)}}
\newcommand{\gammo}[1]{\ensuremath{\Gamma(#1)}}

\newcommand{\Vus}{\ensuremath{V_\mathrm{us}}}
\newcommand{\Vud}{\ensuremath{V_\mathrm{ud}}}
\newcommand{\Vub}{\ensuremath{V_\mathrm{ub}}}
\newcommand{\Vusfo} {\ensuremath{|V_{\rm us} f_{\rm +}(0)|}}
\newcommand{\fo}    {\ensuremath{f_{\rm +}(0)}}

\newcommand{\Fig}{Fig.\,}

\newcommand{\about}{\ensuremath{\sim \,}}

\begin{document}
\vspace*{4cm}
\title{KAON PHYSICS AT KLOE}

\author{THE KLOE COLLABORATION\footnote {
    {F.~Ambrosino},
    {A.~Antonelli},
    {M.~Antonelli},
    {C.~Bacci},
    {P.~Beltrame},
    {G.~Bencivenni},
    {S.~Bertolucci},
    {C.~Bini},
    {C.~Bloise},
    {V.~Bocci},
    {F.~Bossi},
    {D.~Bowring},
    {P.~Branchini},
    {R.~Caloi},
    {P.~Campana},
    {G.~Capon},
    {T.~Capussela},
    {F.~Ceradini},
    {S.~Chi},
    {G.~Chiefari},
    {P.~Ciambrone},
    {S.~Conetti},
    {E.~De~Lucia},
    {A.~De~Santis},
    {P.~De~Simone},
    {G.~De~Zorzi},
    {S.~Dell'Agnello},
    {A.~Denig},
    {A.~Di~Domenico},
    {C.~Di~Donato},
    {S.~Di~Falco},
    {B.~Di~Micco},
    {A.~Doria},
    {M.~Dreucci},
    {G.~Felici},
    {A.~Ferrari},
    {M.~L.~Ferrer},
    {G.~Finocchiaro},
    {S.~Fiore},
    {C.~Forti},
    {P.~Franzini},
    {C.~Gatti},
    {P.~Gauzzi},
    {S.~Giovannella},
    {E.~Gorini},
    {E.~Graziani},
    {M.~Incagli},
    {W.~Kluge},
    {V.~Kulikov},
    {F.~Lacava},
    {G.~Lanfranchi},
    {J.~Lee-Franzini},
    {D.~Leone},
    {M.~Martini},
    {P.~Massarotti},
    {W.~Mei},
    {S.~Meola},
    {S.~Miscetti},
    {M.~Moulson},
    {S.~M\"uller},
    {F.~Murtas},
    {M.~Napolitano},
    {F.~Nguyen},
    {M.~Palutan},
    {E.~Pasqualucci},
    {A.~Passeri},
    {V.~Patera},
    {F.~Perfetto},
    {L.~Pontecorvo},
    {M.~Primavera},
    {P.~Santangelo},
    {E.~Santovetti},
    {G.~Saracino},
    {B.~Sciascia},
    {A.~Sciubba},
    {F.~Scuri},
    {I.~Sfiligoi},
    {T.~Spadaro},
    {M.~Testa},
    {L.~Tortora},
    {P.~Valente},
    {B.~Valeriani},
    {G.~Venanzoni},
    {S.~Veneziano},
    {A.~Ventura},
    {R.Versaci},
    {G.~Xu}
}}
\address{presented by MATTEO PALUTAN\\INFN, Laboratori Nazionali di Frascati, 
via E.~Fermi 40, 00044 - Frascati, Italy }

\maketitle
\abstracts{The most precise determination of \Vus\ comes from semileptonic kaon decays. We have measured with the KLOE detector
at \Dafne, the Frascati \ff, all the experimental inputs to \Vus\ for both neutral and charged kaons. Using our results
we extract the value of \Vus\ with \about 0.9\% fractional error, which is entirely dominated by the theoretical error on the
knowledge of the normalization of the decay form factors, \fo.
A new determination of the ratio \Vus/\Vud\ is also presented, based on our precise measurement of the absolute branching ratio 
for the decay \Dkmudue{}, combined with recent lattice results for the ratio $f_K/f_{\pi}$.
New results on CPT symmetry have also been achieved, which are based on the first measurement of the charge asymmetry for \DKSeIII\ decay.}

\section{Introduction}

The CKM matrix element \Vus\ can be extracted from the measurement  
of the semileptonic decay widths ($\Gamma = {\rm BR}/\tau$) and
the most precise test of unitarity of the CKM matrix is performed
from the first-row constraint: $1-\Delta\!=\!|\Vud|^{2}\!+\!|\Vus|^{2}\!+\!|\Vub|^{2}$.
Using \Vud\ from nuclear beta decays, a test of the expectation $\Delta\!=\!0$ with a 
precision of one part per mil can be performed, \Vub\ contributing only at the level
of $10^{-5}$.  With the KLOE detector 
we can measure all
experimental inputs to \Vus: branching ratios, lifetimes, and decay form
factors.

In the \Dafne\ \Pep\Pem~collider, beams collide at a center-of-mass 
energy $W \sim M(\Pphi)$.
 Since 2001, KLOE has collected an
integrated luminosity of \about2500\Lpb.
Results presented below are based on 2001-2002 data, corresponding to an integrated luminosity of 
\about450\Lpb.
The KLOE detector consists of a large cylindrical drift chamber surrounded by a 
lead/scin\-til\-la\-ting-fiber electromagnetic calorimeter. A superconducting coil around 
the calorimeter provides a 0.52~T field. The drift chamber,~\cite{kloe:dc} 
is 4\m\ in diameter and 3.3\m\ long. 
The momentum resolution is $\sigma(p_{T})/p_{T} \sim 0.4\%$. Two track vertices
are reconstructed with a spatial resolution of $\sim$ 3\mm. The calorimeter,~\cite{kloe:emc} 
composed of a barrel and two endcaps, covers 98\% of the solid angle.
Energy and time resolution are $\sigma(E)/E = 5.7\%/\sqrt{E(\GeV)}$ and
$\sigma(t) = 57~\ps/\sqrt{E(\GeV)} \oplus 100~\ps$.
The KLOE trigger,~\cite{kloe:trg} uses calorimeter and drift chamber information.
For the present analysis only the calorimeter signals are used. Two energy deposits
above threshold, $E>50$~\MeV\ for the barrel and $E>150$~\MeV\ for the endcaps, are required.

The \Pphi\ meson decays mainly into kaons: 49\% to \kp\km and 34\% to \ks\kl\ pairs.
We can thus tag \kl, \ks, \kp, and \km\ decays by detecting respectively 
\ks, \kl, \km, and \kp\ decays on the opposite hemisphere. 
The tagging technique is a unique opportunity of a \ff, and provides pure and almost monochromatic ``beams'' of 
kaons, thus allowing to measure the absolute branching ratios. 
The \ks\ beam is tagged using events 
with a \kl\ interacting in the calorimeter (\kl-crash).
\kl-mesons are tagged detecting \KSpippim\ decays. 
Charged kaons are tagged using two-body decays, \Dkmudue{\pm} and \Dkpidue{\pm}. 
For all of the cases it is possible to
precisely measure the tagged kaon momentum from the knowledge of the \Pphi\ and the tagging kaon momenta. 

\section{\kl\ decays}

\subsection{Major \kl\ decays and lifetime}

As already stated, a pure sample of \kl\ decays is selected by
the identification of \KSpippim\ decays. Events include \kl\
which decay to any possible final state in the detector volume,
interact in the calorimeter or escape the detector. Starting from
this sample, the \kl\ branching ratios are evaluated by counting
the number of decays to each channel in the fiducial
volume (FV) and correcting for the geometrical acceptance, 
the reconstruction efficiency, and the background contamination.

\begin{figure}[htbp!]
  \begin{center}
    \includegraphics[totalheight=6.2cm]{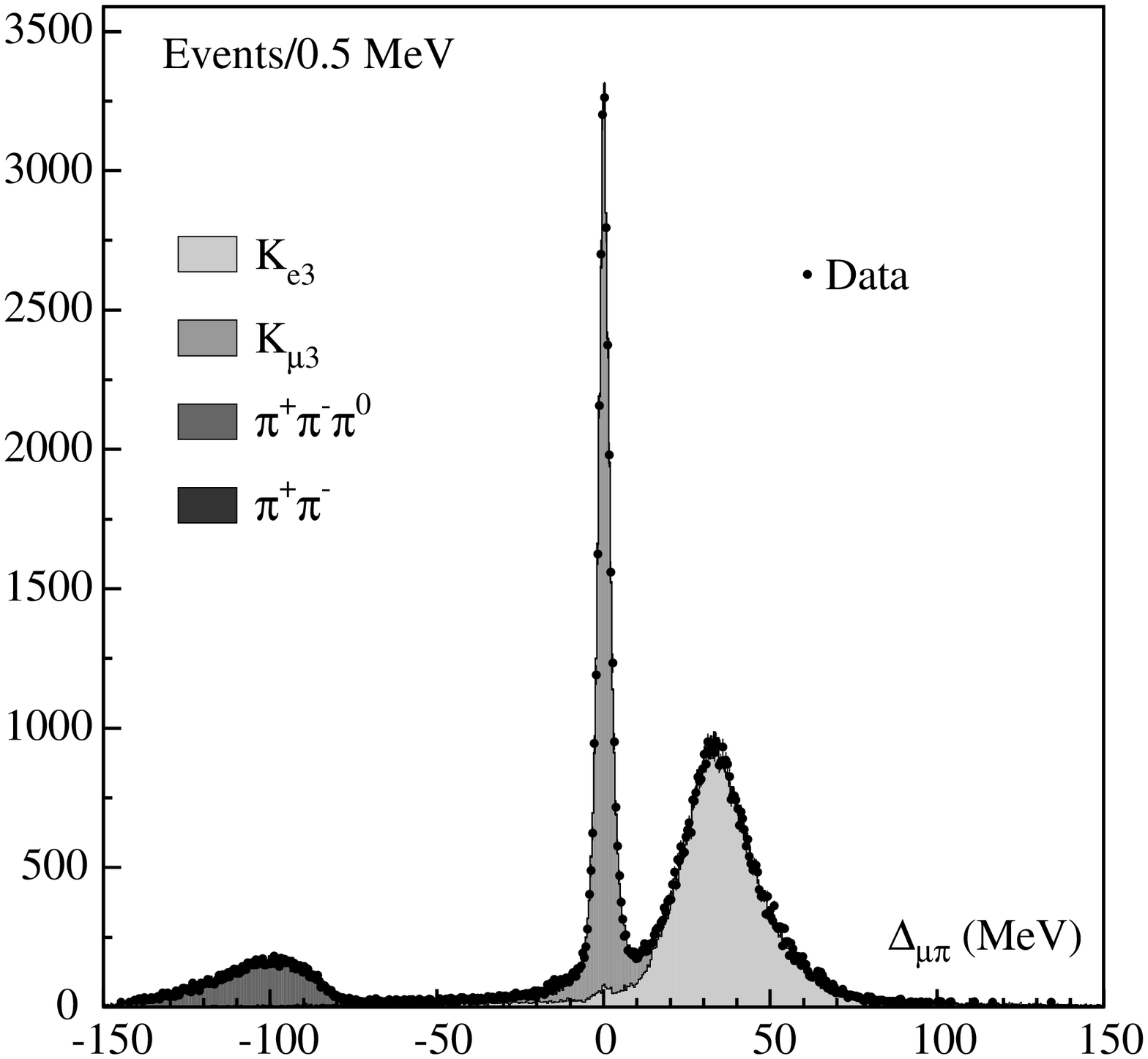}
    \includegraphics[totalheight=6.2cm]{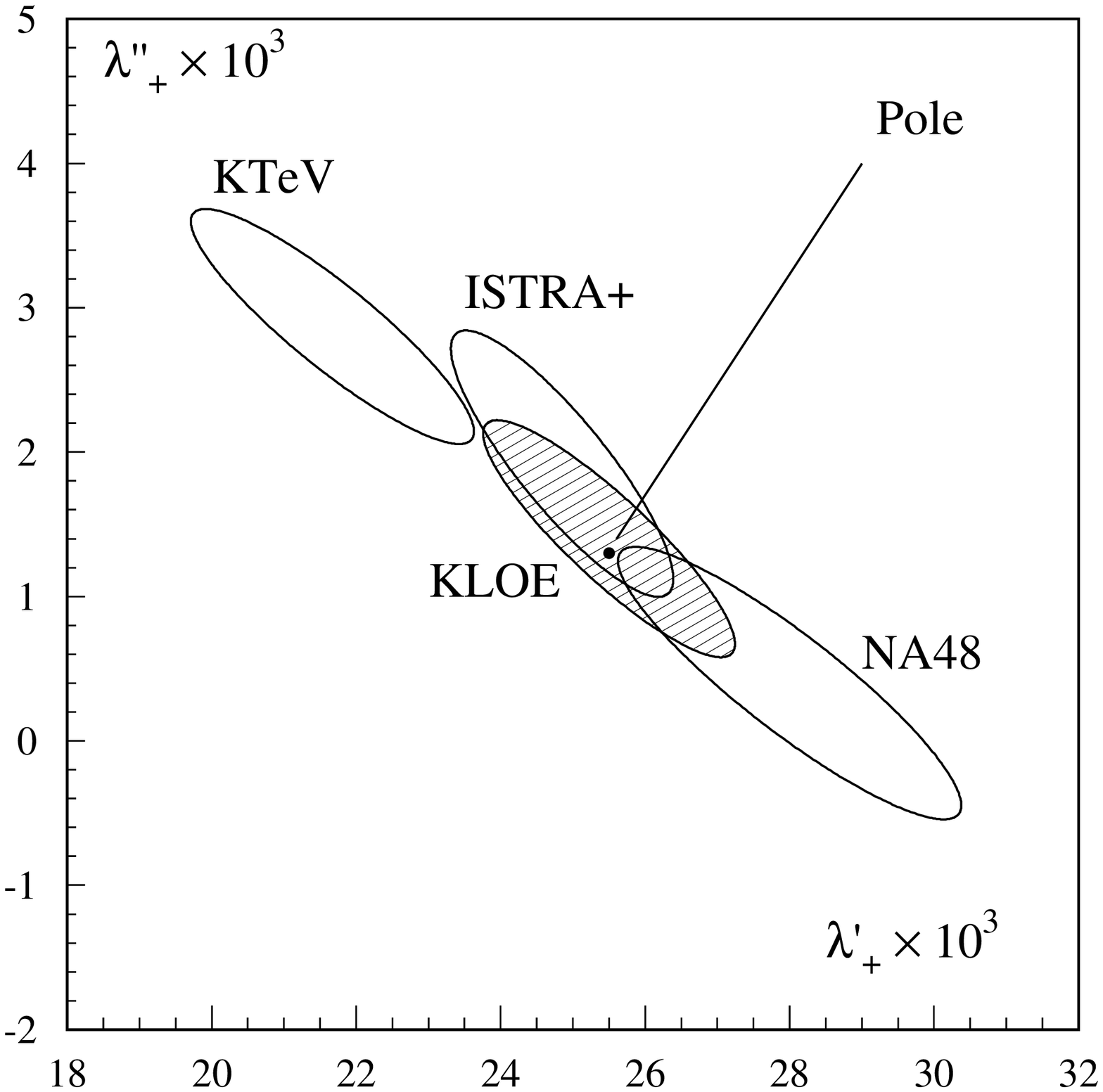}
    \caption{Left panel: distribution of $\Delta_{\mu \pi}$ for a data subsample, with fit to MC 
distributions for different decay channels. Right panel: KLOE results on semileptonic form factor 
slopes compared with other recent measurements; the black dot represents the values of  
$\lambda_{+}^{\prime}$ and $\lambda_{+}^{\prime\prime}$ obtained from the
Taylor expansion of the pole parametrization.}
    \label{figklmost}
  \end{center}
\end{figure}

\kl\ charged decays are identified by selecting a decay vertex in
the FV matching the expected
\kl\ flight direction, as defined by the tag.
In order to discriminate among the different \kl\ charged modes the variable
 $\Delta_{\mu \pi} = |p_{miss}-E_{miss}|$ is used, where $p_{miss}$ and
$E_{miss}$ are the missing momentum and missing energy at the \kl\
vertex, evaluated by assigning one track the pion mass and the
other one the muon mass (left panel of \Fig\ref{figklmost}). Signal counting is thus
achieved by fitting the  $\Delta_{\mu \pi}$ spectrum with a linear combination
 of four Monte Carlo shapes ($\kl \rightarrow \pi e \nu, \pi \mu \nu,
\pi^+\pi^-\pi^0, \pi^+\pi^-$). 
The shape of $\Delta_{\mu \pi}$ spectrum 
is sensitive to
the radiative corrections to \kl\ decay processes. The KLOE Monte
Carlo includes very accurate generators for all of the radiative
channels, which has been studied and implemented mostly in 2004.~\cite{radiativearticle}.

To count \DKLpiopiopio\ events decaying in the FV, we exploit the
time-of-flight capability of the calorimeter to reconstruct the
neutral vertex position. Such a vertex is assumed to be along the
\kl\ line of flight.  The arrival time of each photon identified in the
calorimeter is thus used to give an independent determination of
the path length of the \kl, $L_{K}$. The final value of $L_{K}$ is obtained
 from a wheighted average of the different measurements coming from all of
the photons.
The analysis of the \DKLpiopiopio\ sample is used not only to count events for
BR measurement, but also to
extract the \kl\ lifetime ($\tau_{\kl}$) from a fit to the proper time
distribution of neutral decay vertices. The KLOE environment is indeed 
very well suited for a direct measurement of $\tau_{\kl}$, because of the
 tagging technique 
which provides the value of the \kl\ momentum and allows to select a pure
sample of \kl\ decays.
The result is $\tau_{\kl}
= 50.92 \pm 0.17 \pm 0.25 $~ns.~\cite{kloe:taul}  

A total of about $13\times 10^6$ tagged events have been used for
the measurement of the \kl\ BR's. Since the geometrical efficiency of the fiducial
 volume
depends on the value of the \kl\ lifetime, the sum of all of the decay modes
 is fixed to unity
(BR's below 1\% from PDG~\cite{PDBook}). This removes the uncertainty due to
$\tau_{\kl}$, while giving at the same time a precise
determination of \kl\ lifetime itself. Results for the main BR's are:~\cite{kloe:brl}

\begin{equation}
  \begin{array}{rcl}
    \BR{\DKLeIII} & = & 0.4007 \pm 0.0005_{\rm stat}\pm
    0.0004_{\rm syst-stat}\pm 0.0014_{\rm syst},
    \\
    \BR{\DKLmuIII} & = & 0.2698 \pm 0.0005_{\rm stat}\pm
    0.0004_{\rm syst-stat}\pm 0.0014_{\rm syst},
    \\
    \BR{\DKLpiopiopio} & = & 0.1997 \pm 0.0003_{\rm stat}\pm
    0.0004_{\rm syst-stat}\pm 0.0019_{\rm syst},
    \\
    \BR{\DKLpippimpio} & = & 0.1263 \pm 0.0004_{\rm stat}\pm
    0.0003_{\rm syst-stat}\pm 0.0011_{\rm syst}.
  \end{array}
\end{equation}

The corresponding lifetime is $\tau_{\kl} = 50.72 \pm 0.11_{\rm
stat}\pm 0.13_{\rm syst-stat}\pm 0.33_{\rm
syst}$~ns, in agreement with the direct measurement
that has been shown before.
The two lifetime measurements are uncorrelated; their average is
$\tau_{\kl} =  50.84 \pm 0.23$\ns,
and has a fractional error of 0.45\%. This represents a factor of two 
improvement with respect to the previous measurements.~\cite{vosburg}

\subsection{Semileptonic form factor slopes}

In semileptonic kaon decays, $K_{L,S} \to \pi^{\pm} \ell^{\mp} \nu$, 
only the vector part of the weak current has a non-vanishing 
matrix element between a kaon and a pion. The vector current is ``almost'' conserved. For a vector interaction, there are no 
$SU(3)$-breaking corrections to first order in the $s$-$d$ mass difference, making calculations of hadronic 
matrix elements more reliable. In the electron mode $K_{L,S} \to \pi^{\pm} e^{\mp} \nu$, only 
the vector form factor $f_{+}(t)$ is involved, since extra terms in the matrix element depend on the electron mass.
This form factor is usually parametrized as
\begin{equation}
f_{+}(t)=f_{+}(0)\left[1+\lambda_{+}^{\prime}\frac{t}{m^{2}_{\pi^{+}}}+\frac{\lambda^{\prime\prime}_{+}}{2}\frac{t^{2}}{m^{4}_{\pi^{+}}} + ...\right],
\label{eq:ff}
\end{equation}
 where $f_{+}(0)$ is evaluated from theory and $t$ is the lepton-pair invariant mass squared.
We used about  2 million of $\DKLeIII$ decays from 2001-2002 data sample to measure the
semileptonic form factor slopes.
 These have been extracted by fitting the spectrum of $t/m^{2}_{\pi^{+}}$ for the  
selected events.  The fit procedure takes into account the efficiency of 
the selection cuts, the resolution effects and the background contamination as a function of $t$.
The results obtained for a quadratic fit are:~\cite{kloe:ff}
\begin{equation}
  \begin{array}{rcl}
    \lambda_{+}^{\prime} & = &  (25.5 \pm 1.5_{\rm stat} \pm 1.0_{\rm syst})\times 10^{-3} \\
    \lambda_{+}^{\prime\prime} & = & (1.4 \pm 0.7_{\rm stat} \pm 0.4_{\rm syst})\times 10^{-3},\\
  \end{array}
\end{equation}
which gives a $\chi^2/ndf = 325/362$ ($P(\chi^2) = 92\%$).

We have also fit the data using a pole parametrization: $f_+(t)/f_+(0) = M^2_V/(M^2_V - t)$. This
assumes the form factor is dominated by the vector $K-\pi$ resonances, the closest being the
$K^{\ast}(892)$. We obtain $M_V = (870 \pm 6_{\rm stat} \pm 7_{\rm syst})\MeV$ ($\chi^2/ndf = 326/363$), 
that confirms the dominance of $K^{\ast}$ meson, although contributions from other $J^P = 1^-$ 
resonant and non-resonant $K\pi$ scattering amplitudes are not negligible.
Note that the quadratic parametrization of the form factor arises naturally as a Taylor 
expansion of the pole model, with the additional contraint 
$\lambda_+^{\prime\prime} = 2\lambda_+^2$. This
is nicely fullfilled by our result, which is competitive with other recent 
measurements~\cite{ktevff:04}$^{,}$~\cite{istra+ff:04} (right panel of \Fig\ref{figklmost}).

\section{\DKSeIII\ decay}

At the \ff\ very large samples of tagged, monochromatic \ks\ mesons are 
available. From the analysis of 2001-2002 data sample, we were able 
to obtain a very pure sample of $\sim 13\,000$ \ks\ semileptonic decay events
 and thus
to accurately measure for the first time the partial decay rates for 
transitions to each charged final state, \BR{\ks\toP\Pep\Ppim\Pnu} and 
\BR{\ks\toP\Pem\Ppip\Pnubar}, as well as the charge asymmetry $A_S$:
\begin{equation}
A_{S}=
\frac{
  \Gamma\left({\mathrm{K_{S}\toP\Ppim\Pep\nu}}\right) -
  \Gamma\left({\mathrm{K_{S}\toP\Ppip\Pem\bar{\nu}}}\right)  } 
     {\Gamma\left({\mathrm{K_{S}\toP\Ppim\Pep\nu}}\right) +
       \Gamma\left({\mathrm{K_{S}\toP\Ppip\Pem\bar{\nu}}}\right) }.
\label{eq:asy}
\end{equation}
The comparison of $A_S$ with the corresponding asymmetry $A_L$ for \kl\ decays
allows precision tests of the $CP$ and $CPT$ symmetries. 
If $CPT$ symmetry is assumed, both \ks\ and \kl\ charge asymmetries are 
expected to be 
equal to $2\,\mathrm{Re}\,\epsilon\!\simeq\!3\!\times\!10^{-3}$.
The difference between the charge asymmetries,
\begin{equation}
A_{S}-A_{L}=4\,\left(\mathrm{Re}\,\delta+\mathrm{Re}\,x_{-}\right),
\label{eq:rexm}
\end{equation}
signals $CPT$ violation either in the mass matrix ($\delta$ term), or in the 
decay amplitudes with $\Delta S\neq\Delta Q$ ($x_{-}$ term).
The sum of the asymmetries, 
\begin{equation}
A_{S}+A_{L}=4\left(\mathrm{Re}\,\epsilon-\mathrm{Re}\,y\right),
\label{eq:rey}
\end{equation}
is related to $CP$ violation in the mass matrix ($\epsilon$ term) and to $CPT$
 violation in the decay amplitude ($y$ term).
The knowledge of both the \kl\ and the \ks\ semileptonic decay branching 
ratios 
and lifetimes allows the validity of the $\Delta S\!=\!\Delta Q$ rule to be tested
through the quantity
\begin{equation}
\mathrm{Re}\,x_{+} = \frac{1}{2}\frac{\Gamma(\DKSeIII)-\Gamma(\DKLeIII)}{\Gamma(\DKSeIII)+\Gamma(\DKLeIII)}.
\label{rex}
\end{equation}
In the SM, $\mathrm{Re}\,x_{+}$ is of the order of $G_{F} m_{\pi}^2 \sim 10^{-7}$,
being due to second order weak transitions. 
Finally, from the semileptonic decays of \ks\ a competitive measurement of \Vus\ can be extracted,
which also profits of a very precise knowledge of the \ks\ lifetime.~\cite{PDBook}

\begin{figure}[htbp!]
  \begin{center}
    \includegraphics[totalheight=6.2cm]{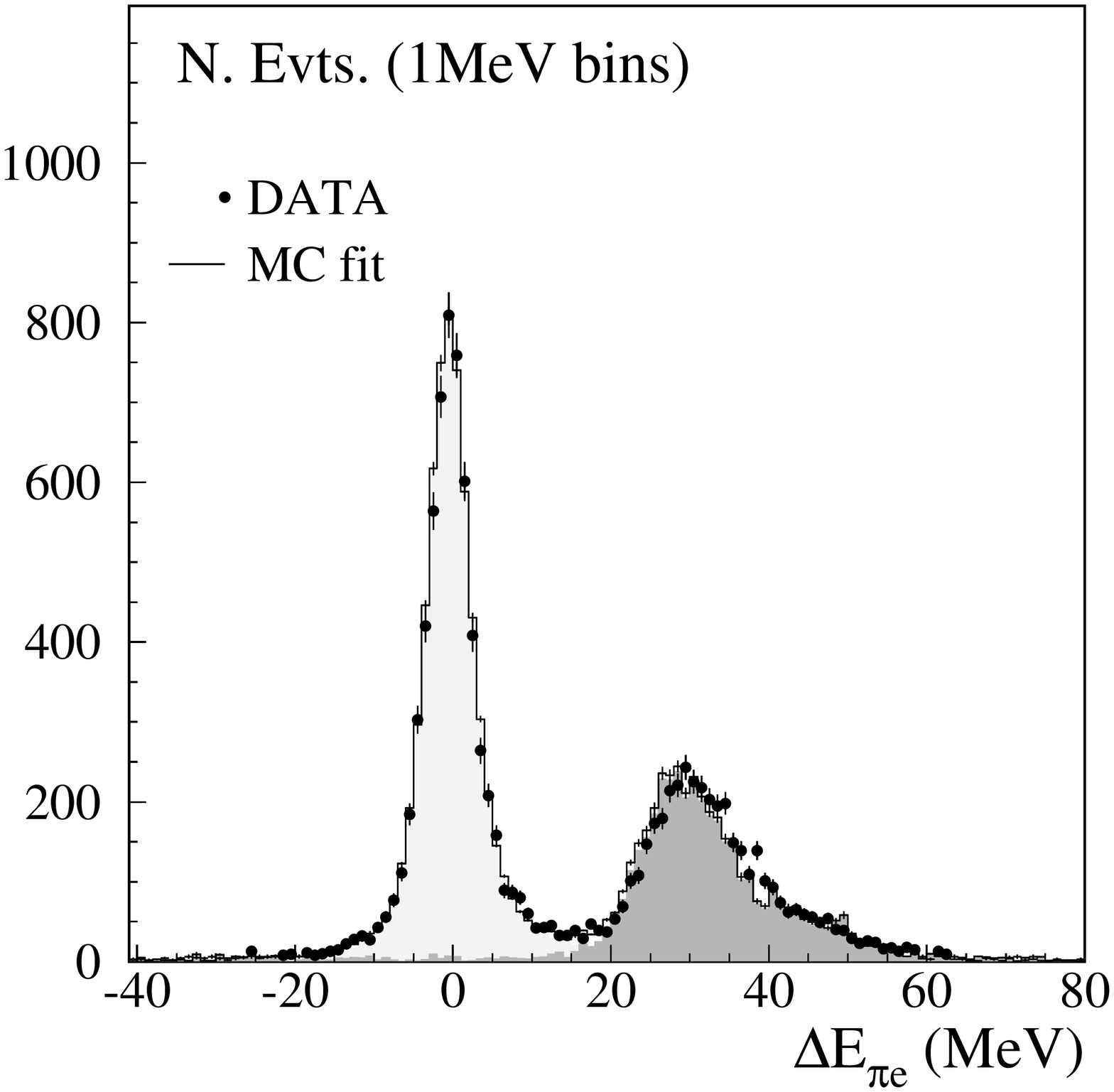}
    \includegraphics[totalheight=6.2cm]{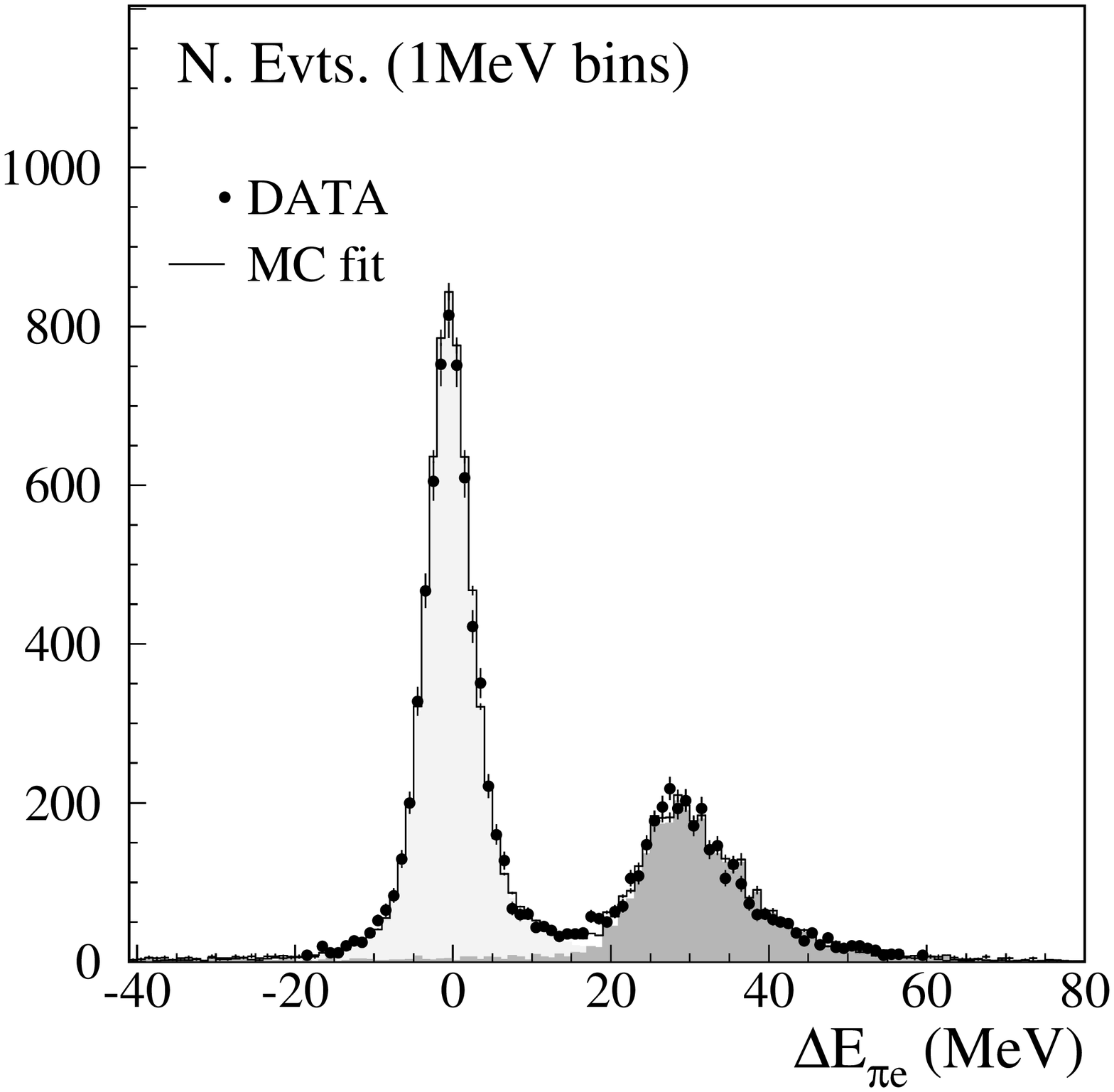}
    \caption{$E_{miss}-p_{miss}$ spectrum for events selected as $\pi^-e^+\nu$ (left panel) and 
$\pi^+e^-\overline{\nu}$ (right panel); filled dots represent data from the entire data set; solid line is
the result of a fit varying the normalization of MC distributions for signal (light gray) and background 
(dark gray),
which are also shown.}
    \label{figksemil}
  \end{center}
\end{figure}

The first measurement of \BR{\DKSeIII} was obtained from KLOE
using data collected in 2000, with a fractional accuracy of 5.4\%.~\cite{Aloisio:2002rq}
The present BR measurement improves
on the total error by a factor of four, to 1.3\%. The \ks\ charge asymmetry has never been measured before. 
Counting of events is performed by fitting the $E_{miss}-p_{miss}$ spectrum with a combination of MC shapes 
for signal and background
(\Fig\ref{figksemil}).  
For each charge state we obtain the ratio 
\BR{\ks\toP\pi^{\mp}e^{\pm}\nu(\overline{\nu})}/\BR{\KSpippim} by normalizing the number 
of signal events to the number of \KSpippim\ events
and correcting for the overall selection 
efficiencies.
In order to evaluate the BR's for the semileptonic modes,
we combine the measured ratios with
the ratio \R\ of the dominant \ks\ decay modes 
which is measured at KLOE.
We obtain the following results:~\cite{kloe:kspen06} 
\begin{equation}
  \begin{array}{rcl}
    \BR{\DKSeIIIeppm} & = & (3.528\pm0.062)\times10^{-4}\\
    \BR{\DKSeIIIempp} & = & (3.517\pm0.058)\times10^{-4}\\
    \BR{\DKSeIII}     & = & (7.046\pm0.091)\times10^{-4}.\\
  \end{array}
\end{equation}
The charge asymmetry of Eq.~(\ref{eq:asy}) is evaluated to be:~\cite{kloe:kspen06}
$A_{S}\!=\! (1.5\pm9.6_{\mathrm{stat}}\pm2.9_{\mathrm{syst}}) \times 10^{-3}$.
From the total BR we test the validity of the $\Delta S=\Delta Q$ rule in $CPT$-conserving 
transitions (Eq.~\ref{rex}). 
In the evaluation of $\Gamma's$ we used $\tau_{\ks} = 0.08958(6)$\ns\ from the PDG~\cite{PDBook} 
and KLOE results for $\tau_{\kl}$ and  \BR{\DKLeIII}.
We obtain $\mathrm{Re}\,x_{+} = \left(-0.5\pm3.6\right)\times 10^{-3}$.~\cite{kloe:kspen06}
The error on this measurement represents an improvement by almost a factor of two with respect to the 
most precise previous result.~\cite{CPLEAR_rex:98}

From the sum and difference of the \kl\ and \ks\ charge asymmetries one can test for possible 
violations of the $CPT$ symmetry, 
either in the decay amplitudes 
or in the mass matrix (Eqs.~\ref{eq:rexm} and \ref{eq:rey}).
Using $A_L = (3.34\pm0.07)\times10^{-3}$,~\cite{PDBook} and $\mathrm{Re}\,\delta=(3.0\pm3.3_{\rm stat}\pm0.3_{\rm syst})\times10^{-4}$,~\cite{CPLEAR_redelta:98} we obtain
$\mathrm{Re}\,x_{-}=\left(-0.8\pm2.5\right) \times 10^{-3}$.~\cite{kloe:kspen06}
 This result improves by a factor of ten respect to the previous best result.~\cite{CPLEAR_redelta:98} 
Using  $\mathrm{Re}\,\epsilon=(1.62\pm0.04)\times10^{-3}$,~\cite{PDBook} we obtain
  $\mathrm{Re}\,y=\left(0.4\pm2.5\right) \times 10^{-3}$,~\cite{kloe:kspen06}
which has precision comparable to that ($3\!\times\!10^{-3}$) obtained from the unitarity relation by CPLEAR.~\cite{CPLEAR:bell}

Finally, we measured the form factor slope
by fitting the ratio of data and MC distributions in $t/m_{\pi+}^{2}$.  
The fit has been performed by using only a linear parametrization of the form factor, since the available 
statistics do not allow to be sensitive to a quadratic one. 
The linear slope of the semileptonic \ks\ form factor is measured to be 
 $ \lambda_{+} = (33.9\pm4.1)\times10^{-3}$.~\cite{kloe:kspen06}
This result is in agreement with the corresponding value for the linear slope of the semileptonic \kl\ 
form-factor.

\section{Charged kaon decays}

\subsection{Semileptonic \kpm\ decays}

\begin{figure}[htbp!]
  \begin{center}
    \includegraphics[totalheight=7.2cm]{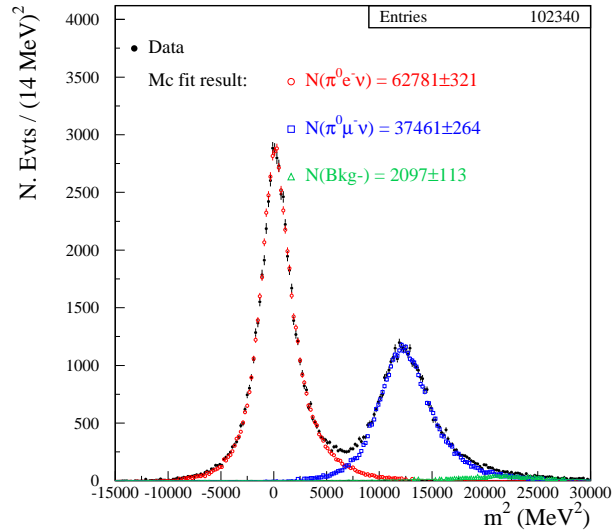}
    \caption{ Lepton mass distribution of the \kltre{-} sample tagged by \kmudue{+} events; 
\ketre{}\ events are recognized as a peak around $m_{lept}^2=0$, while \kmutre{}\ events peak
around  $m_{lept}^2 \sim m_{\mu}^2$. }
    \label{spectra1} 
  \end{center}
\end{figure}

The measurement of the branching ratios for the \kpm\ semileptonic decays
is performed using four data samples defined by different decay modes and charge states of the 
tagging kaon: \kmudue{+}, \kpidue{+}, \kmudue{-}, and \kpidue{-}.
This redundancy allows to keep under control the systematic effects due to the tag selection.
Kaons are identified as tracks with momentum $70<{\rm p}<130$\MeV,
originating from the collision point. The kaon decay vertex must be
within a fiducial volume (FV) defined as 
a cylinder of radius $40<{\rm r}<150$\cm, centered at the collision
point, coaxial with the beams. The decay track, extrapolated to the
calorimeter, must point to an appropriate energy deposit.
\kmudue{} (\kpidue{}) decays are selected by applying a 3$\sigma$ cuts around 
the muon (pion) momentum calculated in the kaon rest frame, according to the
proper mass hypothesis.
For the \kpidue{\pm} tag, identification of the \Ppio\ from the vertex is also required. Finally,
to reduce the dependence of the tag selection efficiency on the decay mode 
of the signal kaon, the tagging decay is required to satisfy the calorimeter trigger by itself.
To select a semileptonic decay on the signal side, a one-prong kaon decay 
vertex must be present in the FV.
The daughter track has to reach the calorimeter and to overlap an energy deposit;
two-body decays are rejected by requiring that its momentum in the kaon
frame, computed assuming the pion mass, is less than 195 \MeV.
The lepton mass ($m_{lept}$) is obtained from the velocity of the lepton computed from the time of flight.
The number of \ketre{} and \kmutre{} decays is then obtained by
fitting the $m_{lept}^2$ distribution (\Fig\ref{spectra1}) to a sum of MC distributions for
the signals and the various background sources.

The BR is evaluated separately for each tag sample, dividing the number of signal counts by the number 
of tag events and correcting for acceptances.
The latter are obtained from MC simulation, and correcting for data-MC 
differences in tracking efficiency and calorimeter clustering.
About 190\,000 \ketre{\pm} and 100\,000 \kmutre{\pm} decays are selected. Preliminary results for the
 BR's are:
\begin{equation}
  \begin{array}{rcl}
    \BR{\ketre{\pm}} & = & (5.047 \pm 0.046_{\rm stat} \pm 0.080_{\rm syst})\% \hspace{0.2cm} \\
    \BR{\kmutre{\pm}} & = & (3.310 \pm 0.040_{\rm stat} \pm 0.070_{\rm syst})\%{\mbox .} \\
  \end{array}
\end{equation}
These values have been averaged over the four different tag samples for each channel. 
The quoted systematic error is still preliminary.

\subsection{\kpm\ lifetime}
The \kpm\ lifetime ($\tau_{\kpm}$) is an experimental input to the determination of \Vus. 
The present fractional uncertainty is about 0.2\%,  corresponding to an uncertainty of 0.1\% for \Vus.
However, there are large discrepancies between results of different experiments.~\cite{PDBook} 
The value of $\tau_{\kpm}$ also affects the geometrical acceptance in BR measurements.
KLOE is measuring $\tau_{\kpm}$ with high-statistics using two different methods: one based on the 
measurement of the decay length 
and the other on the decay time of the kaons.
Comparison between the two methods will allow to decrease the systematic error.
The preliminary result from the decay length measurement is
$\tau_{\kpm} = 12.336 \pm 0.044_{\rm stat} \pm 0.065_{\rm syst}$\ns.

\section{Determination of \Vus}

 \Vus\ is proportional to the square root of the
partial width of semileptonic kaon decays , and can be parametrized, for neutral kaon decays,
as:
\begin{equation}
 \Vusfo = \left[
\frac{128 \pi^3 \Gamma }{G_{\mu}^2 M_{K}^5
S_{\mathrm{ew}}I(\lambda_+^\prime,\lambda^{\prime\prime}_+,\lambda_0,0) }
\right]^{1/2} \frac{1}{1 + \delta_{{\mathrm{em}}}}, \label{vus}
\end{equation}
where \fo\ is the vector form factor at zero
momentum transfer and $I(\lambda_+^{\prime},\lambda^{\prime\prime}_+,\lambda_0,0)$
is the result of the phase space integration after factorizing out
\fo. In the above expression, long-distance
radiative corrections for both the form factor \fo\
and the phase space integral have been factorized out and are
included in the parameter $\delta_{{\mathrm{em}}}$, which amounts
to $5-8\times 10^{-3}$ for $K_{e3}$ and $K_{\mu 3}$
respectively.~\cite{Cirigliano:2004pv}$^{,}$~\cite{Andre:2004tk}
The short-distance electroweak corrections are included in the
parameter $S_{\mathrm{ew}}=1.0232$;~\cite{Sirlin:82} $\lambda_+^{\prime}$
and $\lambda^{\prime\prime}_+$ are the quadratic slopes of the vector
form factor, and $\lambda_0$ is the slope of the scalar form factor. 

The BR's of the semileptonic \kl\ decays (and $\tau_{\kl}$), together with the measurement of 
\BR{\DKSeIII}  and the preliminary results
on the semileptonic \kpm\ decays, give five independent determinations of the 
observable \Vusfo.
The average of KLOE results is $\Vusfo\ = 0.2166 \pm 0.0005$, with  $\chi^2/ndf = 1.9/4$.
The fractional uncertainty is about 0.25\%.
A precise estimate of $\fo = 0.961 \pm 0.008$ was first given in 1984,~\cite{Leutwyler:1984je} and it
has been recently confirmed by 
 lattice calculations.~\cite{Becirevic:04} Using the previous value for \fo\ and the KLOE result for \Vusfo\
we find $\Vus\ = 0.2254 \pm 0.0020$. This has to be compared with the value obtained from unitarity
contraint: $\Vus = \sqrt{1-\Vud^2}=0.2275 \pm 0.0012$.~\cite{marciano:06}    

\begin{figure}[htbp!]
   \begin{center}
    \includegraphics[totalheight=7.2cm]{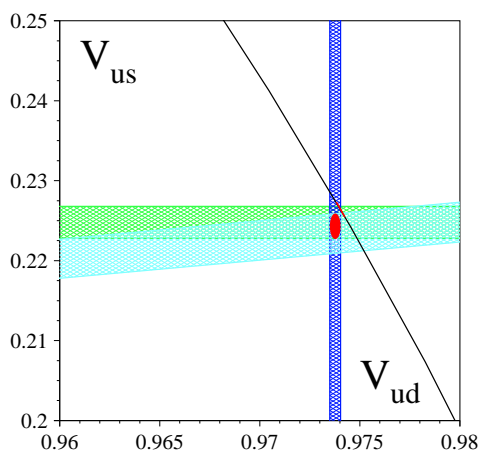}
    \caption{Pictorial view of \Vus, \Vud and \Vus/\Vud\ measurements in the \Vus\ vs \Vud\ plane.
      The small ellipse represents the fit result for \Vus\ and \Vud; the unitariry contraint is represented
      by the black line.}
    \label{fig:vusvud} 
   \end{center}
 \end{figure}

A complementary approach to the previous determination of \Vus\ is the extraction of $|\Vus|/|\Vud|$ 
from the measurement of the ratio of kaon and pion leptonic decay rates, which is parametrized by the equation
\begin{equation}
\frac{\gammo{\Dkmudue{}}}{\gammo{\Dpimudue}{}} = \frac{|\Vus|^2}{|\Vud|^2} 
\frac {  m_K \left( 1 - \frac{m_{\mu}^2}{m_K^2} \right)^2 }{  m_{\pi} \left( 1 - \frac{m_{\mu}^2}{m_{\pi}^2} \right)^2 }
\frac{f_{K}^2}{f_{\pi}^2}
\frac{1 + \frac{\alpha}{\pi} C_K }{1 + \frac{\alpha}{\pi} C_{\pi} }, 
\end{equation}
where $f_{K}$ and $f_{\pi}$ are the kaon and the pion decay contants, respectively; $C_{K}$ and $C_{\pi}$ parametrize the radiative 
electroweak corrections. 
KLOE has recently measured the absolute \kmudue{+}\ branching ratio, with inclusion of photon radiation in the final state:
  \BR{\Dkmudue{+}}=0.6366$\pm$0.0009$_{\rm stat}$ $\pm$0.0015$_{\rm syst}$.~\cite{kloe:kmu2} Using a recent lattice calculation of the
ratio $f_K/f_{\pi}$ of pseudoscalar meson decay constants,~\cite{milc} we find  $\Vus/\Vud = 0.2294 \pm 0.0026$. The fractional error
is \about 1\%, and is dominated by the uncertainty in the $f_K/f_{\pi}$ ratio.
 The previous results on \Vus\ and \Vus/\Vud,  together with the present knowledge of \Vud\ are shown in 
\Fig\ref{fig:vusvud};  
a black line represents the unitarity constraint. A combined fit of the experimental results gives $\Vus = 0.2246\pm0.0016$.
In the same \Vus-\Vud\ plane, the experimental data match the unitarity constraint with P($\chi^2$)\about 0.23. 

\end{document}